\documentclass[aps,pra,showpacs,amsmath,amssymb,twocolumn,nofootinbib,superscriptaddress]{revtex4-1}

\usepackage[pdftex]{graphicx}
\usepackage{mathrsfs,amsmath,amssymb,amsthm}
\usepackage[colorlinks=true,
            linkcolor=red,
            urlcolor=blue,
            citecolor=blue]{hyperref}

\newcommand{\bra}[1]{\langle #1|}
\newcommand{\ket}[1]{|#1\rangle}
\newcommand{\expt}[1]{\langle #1 \rangle}

\newcommand{\dagop}[1]{\hat{#1}^\dag}

\newcommand{\twid}[1]{\overset{\sim}{#1}}

\newcommand{\bo}[1]{\boldsymbol{#1}}
\newcommand{\tr}[1]{\mathrm{Tr}\{ #1 \}}

\begin{document}

\bibliographystyle{apsrev}

\title{Quantized Nonlinear Gaussian-Beam Dynamics---\\
Tailoring Multimode Squeezed-Light Generation}

\author{R. Nicholas Lanning}
\email[rlanni1@lsu.edu]{}
\author{Zhihao Xiao}
\affiliation{Hearne Institute for Theoretical Physics and Department of Physics $\&$ Astronomy, Louisiana State
University, Baton Rouge, Louisiana 70803, USA}
\author{Mi Zhang}
\author{Irina Novikova}
\author{Eugeniy E. Mikhailov}
\affiliation{Department of Physics, College of William $\&$ Mary, Williamsburg, Virginia 23187, USA}
\author{Jonathan P. Dowling}
\affiliation{Hearne Institute for Theoretical Physics and Department of Physics $\&$ Astronomy, Louisiana State University, Baton Rouge, Louisiana 70803, USA}
\affiliation{ECNU Institute of Physics at NYU Shanghai, Shanghai 200062, China}
\affiliation{CAS-Alibaba Quantum Computing Laboratory, USTC, Shanghai 201315, China}
\affiliation{National Institute of Information and Communications Technology, Tokyo 184-8795, Japan}
\date{\today}

\begin{abstract}
We present a general, second quantization procedure for multi transverse-spatial-mode Gaussian-beam dynamics in nonlinear interactions.
Previous treatments have focused on the spectral density and angular distribution of spatial modes.
Here we go a layer deeper by investigating the complex transverse spatial mode in each angular spatial mode.
Furthermore, to implement the theory, we simulate four-wave mixing and parametric down-conversion schemes, showing how one can elucidate and tailor the underlying multi transverse-spatial-mode structure along with it's quantum properties. 
\end{abstract}
		
% insert suggested PACS numbers in braces on next line
% insert suggested keywords - APS authors don't need to do this
%\keywords{}

\maketitle

\section{Introduction}
Nonlinear optical devices producing single photons, multiple photons with nonclassical correlations, and quantum fields with suppressed noise are the cornerstone of many advanced quantum technologies.
For example, most high-performance schemes for quantum imaging \cite{kolobov2007quantum, tsang2009quantum, giovannetti2009sub, brida2010experimental}, metrology \cite{huver2008entangled, anisimov2010quantum, giovannetti2011advances, motes2015linear}, optomechanics \cite{purdy2013strong, singh2016stable, cripe2018radiation}, cryptography \cite{jennewein2000quantum, scarani2009security, pirandola2015high}, and information \cite{wilde2013quantum, weedbrook2012gaussian}, rely on one or more these resources. 
The pursuit of ever more performance has demanded a full understanding of the spatial distribution and mode composition of quantum light. 

Much progress has been made on several fronts. 
An increasing number of studies have been performed which investigate the spatial distribution and transverse-spatial-mode (TSM) structure of squeezed light \citep{Gerry_and_Knight_book} (that is, light with suppressed quantum noise).
This has prominently been done for squeezed light generated in atomic vapors via the polarization self-rotation effect (PSR) \cite{matsko2002vacuum, mikhailov2008low, barreiro2011polarization, agha2010generation} and four-wave mixing (FWM) \cite{reid1985generation, slusher1985observation, kumar1994degenerate, mccormick2007strong} processes. 
The squeezed vacuum generated via PSR has been studied experimentally using spatial masks \cite{zhang2016spatial}  and optimized for various optical depths \cite{zhang2017multipass}, in each case elucidating more about the spatial-mode structure of the noise suppression. 
Squeezing generated via FWM is typically harnessed as two-mode squeezing and spatial correlations between the twin beams have also been studied in detail \cite{boyer2008generation, corzo2011multi, marino2012extracting, holtfrerich2016control}.
However, FWM has also been used to create a single squeezed beam, which is quadrature squeezed in a large number of spatial modes within the beam \cite{embrey2015observation}.

On the other hand, the problem of creating single photons and entangled photon pairs typically involves the parametric down-conversion (PDC) process \citep{Boyd_book, Gerry_and_Knight_book}.
Progress on this front has focused on characterizing the spectral properties and angular distribution of photons in the PDC twin beam \cite{castelletto2005spatial, brida2010detection, walborn2010spatial, spasibko2012spectral}.
One popular approach for studying the spatial mode content is the Schmidt mode analysis \cite{sharapova2015schmidt}.
As an extension, these models are used to calculate the biphoton rate, coupling efficiency, and heralding efficiency \cite{castelletto2004on, guilbert2015enhancing, schneeloch2016introduction}.
In all of these studies, there has remained a crucial area to be investigated.
In each spatial-angular mode of the output beams of FWM and PDC, there are actually many TSMs, which potentially have quantum correlations. 

In this paper, we present a theory which predicts the TSM structure in each spatial-angular mode of FWM and PDC beams. 
We use this knowledge to predict the variance, covariance, and relative coupling strength between the modes \cite{ma1990multimode}. 
Furthermore, we identify the eigenmodes of the interaction, and use these to show how to enhance the noise suppression of the system \cite{bennink2002improved}.  
Also, we show that, under certain conditions, they can be used as a basis to represent the interaction. 
To demonstrate the theory, we simulate several interactions, including PSR, FWM, and PDC. 
In each case, we focus on exposing the underlying TSM structure and suggest ways to tailor it depending on the intended use of the quantum resources. 
Here, we focus on squeezing and heralding, and entanglement properties will be investigated in a future work.
The simulations herein illustrate the utility of our theory, which is a convenient and powerful optimization tool.
It can be used to show how the interplay of experimental beam parameters can influence the quantum properties of the beam(s) generated during the interaction. 
For example, the mode structure of the input beams, along with their waists and focal positions, can all have significant influence on the quantum properties.
Coupling to the generated beams can also be a formidable problem.
Thus, it allows one to investigate which beam parameters should be targeted to enhance the quantum resources.

\section{Theory}
\subsection{Hamiltonian and Unitary Evolution}
Any third-order nonlinear interaction can be described in terms of the quantum fields $\hat{E}$ by a Hamiltonian of the form
\begin{equation}\label{eq:HAM_1}
	\begin{split}
H \propto \int d\textbf{r}^3 \chi^{(3)}(\textbf{r}) \, \bigg( \hat{E}^{(-)}_{d1}(\textbf{r},t) \, \hat{E}^{(-)}_{d2}(\textbf{r},t) \\
\times \: \hat{E}^{(+)}_{s}(\textbf{r},t) \, \hat{E}^{(+)}_{i}(\textbf{r},t) - \mathrm{H.c.} \bigg),
	\end{split}
\end{equation}
where $\chi^{(3)}(\textbf{r})$ is the third-order nonlinearity, $\hat{E}^{(\pm)}$ corresponds to the positive and negative frequency components of the field, and $d1,d2,s,i$ corresponds to drive (pump), signal, and idler fields. We follow tradition in labeling the non-driving modes signal and idler (also referred to as target modes in literature).

We elect to consider input beams with cylindrical symmetry, but we note that the following calculation can certainly be done in other coordinate systems. 
In cylindrical coordinates, the homogeneous paraxial wave equation gives rise to the Laguerre-Gauss (LG) family of solutions \cite{Siegman_book}:
\begin{equation} \label{eq:LG_MODES}
 \begin{split}
  	u_{\ell ,p}(\vec{r})=& \dfrac{C_{\ell ,p}}{w(z)} e^{- \frac{r^{2}}{w(z)^{2}}}e^ { -\frac{ikr^{2}z}{2(z^{2}+z_{R}^{2})} }  \big( \dfrac{\sqrt{2}r}{w(z)} \big)^{|\ell |} \\
	\times&  \: L_{p}^{|\ell |}
	\big( \dfrac{2r^{2}}{w(z)^{2}} \big)  e^{i \ell \phi} e^{i(2p+|\ell |+1)\arctan(z/z_{R})}, 
 \end{split}
\end{equation}
where $\ell$ is the azimuthal index, $p$ is the radial index for each mode, $ C_{\ell ,p}=\sqrt{2p! / \pi(|\ell |+p)!}$ is a normalization constant, $w_{0}$ is the beam waist, $ w(z)=w_{0}\sqrt{1+(z/z_{R})^2}$ is the width function of the beam, $L_{p}^{|\ell |}$ are the generalized Laguerre polynomials, $z_{R}=\pi w_{0}^{2}/\lambda$ is the Rayleigh range, and $k=2\pi/\lambda$ is the wave
number.

To retain generality, we will assume that the pump beam modes are known but that the signal and idler modes have vacuum inputs.
Therefore, we must allow for a large number of spatial 
 signal and idler fields.
Thus, we let 
\begin{equation} \label{eq:S_I}
\begin{split}
\hat{E}^{(+)}_{s}(\textbf{r},t) &= \sum_{\ell,p} u_{\ell,p}(\textbf{r}) \, \hat{a}_{\ell,p} \, e^{i(\textbf{k}_s \cdot\textbf{r}-  \omega_s t)}\\
\hat{E}^{(+)}_{i}(\textbf{r},t) &= \sum_{m,q} u_{m,q}(\textbf{r}) \, \hat{b}_{m,q} \, e^{i(\textbf{k}_i \cdot\textbf{r}-  \omega_i t)}.
\end{split}
\end{equation}
In reality, an infinite number of modes is not necessary and the sum can be truncated to a total of $N$ modes.
To determine the mode structure relevant to the interaction, one can use our semiclassical-beam theory \cite{lanning2017gaussian}.
Furthermore, if a vacuum mode is replaced with a seed beam, then our semiclassical theory can predict the beam evolution, which can be included here.  
The pump beams, on the other hand, are treated classically and have a well-known structure, and without loss of generality, we can choose them to be Gaussian beams of the form
\begin{equation}
	\begin{split}
\hat{E}^{(+)}_{d1}(\textbf{r},t) &= A_{d1} \, u_{0,0}(\textbf{r}) \, \hat{d_1}  \, e^{i(\textbf{k}_{d1} \cdot\textbf{r}-  \omega_{d1} t)}\\
\hat{E}^{(+)}_{d2}(\textbf{r},t) &= A_{d2} \, u_{0,0}(\textbf{r}) \, \hat{d_2}  \, e^{i(\textbf{k}_{d2} \cdot\textbf{r}-  \omega_{d2} t)},
	\end{split}
\end{equation} 
where $A_{d1}$ and $A_{d2}$ are complex amplitudes.
In practice, one can plug in whatever mode structure is present in the pump beam(s).
Next, we make the parametric approximation and drop the operator character of the pump fields, transforming Eq.~(\ref{eq:HAM_1}) into 
\begin{equation}\label{eq:HAM_2}
	\begin{split}
\hat{H} =& \, \kappa  \int d\textbf{r}^3 \sum_{\ell p m q} \bigg( \chi^{(3)*}_{\ell,p;m,q} \hat{a}_{\ell,p} \,\hat{b}_{m,q} - \chi^{(3)}_{\ell,p;m,q} \hat{a}_{\ell,p}^{\dagger} \,\hat{b}^{\dagger}_{m,q} \bigg),
	\end{split}
\end{equation}
where $\kappa$ is a coupling constant, the effective susceptibility is $\chi^{(3)}_{\ell,p;m,q} \equiv C \, \chi^{(3)}(\textbf{r}) \,  A_{d1} A_{d2} \, u^2_{0,0}(\textbf{r})\,u_{\ell,p}^*(\textbf{r})\,u_{m,q}^*(\textbf{r})  $, $C$ is a normalization constant, and we have assumed phase matching allowing us to drop the exponential factor.

\begin{figure}[!t]
	\includegraphics[width=\columnwidth]{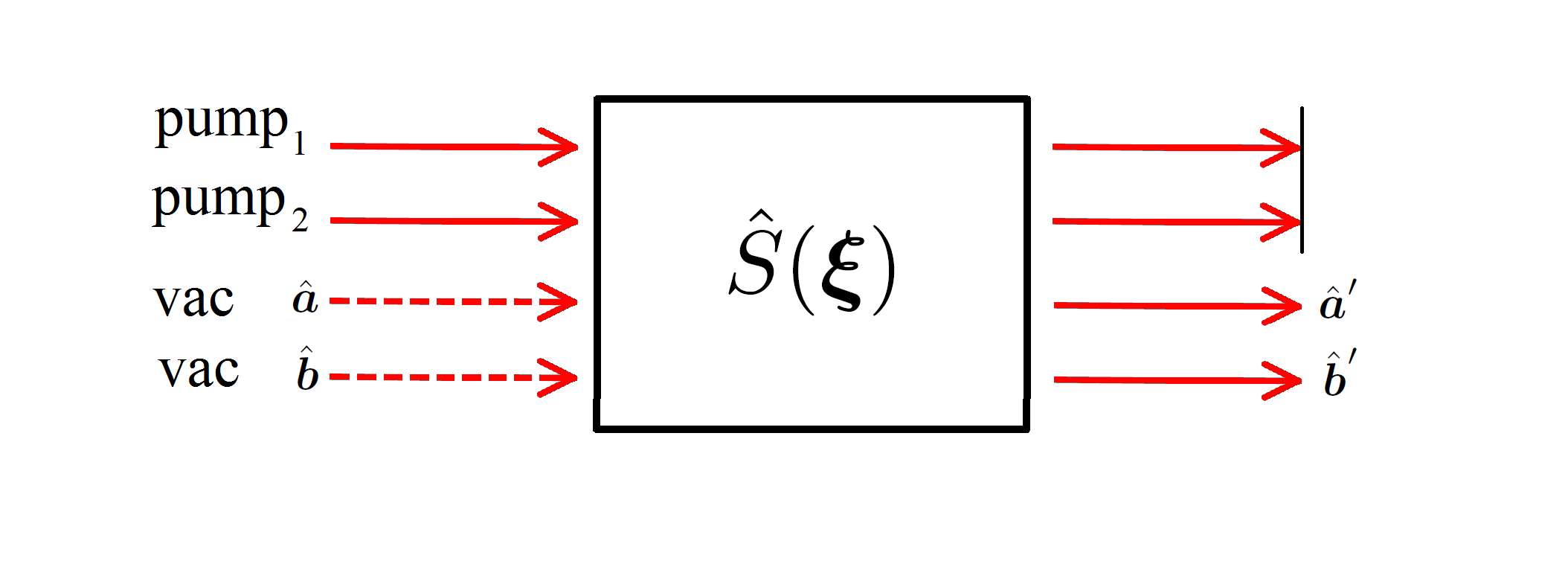}
	\caption{\label{fig:General_Schematic} A general unitary evolution generated in a $\chi^{(3)}$ material. In each spatial mode, there are actually many transverse spatial modes interacting not only within the mode, but potentially with many other modes in the adjacent spacial modes. Our theory can accommodate single- or dual-pump modes and single- or dual-vacuum input modes. Furthermore, a vacuum mode, for example, $\hat{\bo{b}}$, could be replaced with a seed beam and evolution could be calculated with our semiclassical-beam theory for nonlinear optics \cite{lanning2017gaussian} and incorporated into this second quantization procedure.        
	}
\end{figure}

Next, we will simplify notation and make calculations more straightforward by turning the double sum in Eq.~(\ref{eq:HAM_2}) into matrix multiplication.
First, we define the vector of operators 
\begin{equation}\label{eq:VECTORS}
	\begin{split}
\hat{\boldsymbol{a}} &\equiv (\hat{a}_{\ell,p} \; \hat{a}_{\ell,p+1} \;... \; \hat{a}_{\ell+1,p} \;\hat{a}_{\ell+1,p+1} \; ...)^{T}\\
\dagop{\boldsymbol{a}} &\equiv (\dagop{a}_{\ell,p} \; \dagop{a}_{\ell,p+1} \;... \; \dagop{a}_{\ell+1,p} \;\dagop{a}_{\ell+1,p+1} \; ...)^{T},
	\end{split}
\end{equation}
where, in this case, $l$ and $p$ are the lowest-order modes in consideration, $(\cdot)^{T}$ indicates the transpose operation, and $\hat{b}$ follows accordingly.
With these vectors in mind, we define a corresponding two-photon amplitude matrix (see Appendix) and rewrite  Eq.~(\ref{eq:HAM_2}) as 
\begin{equation}\label{eq:HAM_3}
\hat{H} = \int d\textbf{r}^3 \, \bigg( \twid{\hat{\boldsymbol{b}}} \, \boldsymbol{\chi}^{\dagger} \, \hat{\boldsymbol{a}} - \twid{\hat{\boldsymbol{a}}^\dagger} \boldsymbol{\chi} \, \hat{\boldsymbol{b}}^\dagger \bigg),
\end{equation}
where we have introduced $\twid{(\cdot)}$ to denote the transpose.
This Hamiltonian leads to the two-mode squeezing operator
\begin{equation}\label{eq:SQUEEZER}
\hat{S}(\boldsymbol{\xi}) \equiv \exp \bigg[ \twid{\hat{\boldsymbol{b}}} \, \boldsymbol{\xi}^{\dagger} \, \hat{\boldsymbol{a}} - \twid{\hat{\boldsymbol{a}}^\dagger} \boldsymbol{\xi} \, \hat{\boldsymbol{b}}^\dagger \bigg],
\end{equation}
where $\boldsymbol{\xi} \equiv \int d\textbf{r}^3 \, \boldsymbol{\chi} \, t $  is the squeezing matrix.
This unitary evolution is depicted in Fig.~\ref{fig:General_Schematic}.

When $\bo{\xi}$ is symmetric, the left polar decomposition gives $ \boldsymbol{\xi} = \boldsymbol{R} \exp[i \boldsymbol{\Theta}] = \exp[i \twid{\boldsymbol{\Theta}}] \twid{\boldsymbol{R}}$, where $\boldsymbol{R}$ and $\boldsymbol{\Theta}$ are Hermitian matrices. 
In general, $\bo{R}$ and $\bo{\Theta}$ do not commute.
However, it follows that for functions $f$ that are expandable in a power series, $f(\bo{R}) e^{i \bo{\Theta}} =  e^{i \bo{\Theta}} f(\twid{\bo{R}}) $ and $f(\bo{R}) e^{i \bo{\Theta}} =  e^{i \twid{\bo{\Theta}}} f(\twid{\bo{R}}) $, where $f$ is even or odd, respectively.
Straightforward yet tedious repetition of the Baker-Campbell-Hausdorff (BCH) relation allows us to find the Bogoliubov transformations for the multimode vectors:
\begin{equation}\label{eq:BOG_T}
	\begin{split}
\hat{S}^{\dag}(\boldsymbol{\xi}) \,\hat{\bo{a}}\, \hat{S}(\boldsymbol{\xi}) &= \cosh(\bo{R})\hat{\bo{a}} - \sinh(\bo{R})e^{i \bo{\Theta}}\hat{\bo{b}}^{\dag}\\
\hat{S}^{\dag}(\boldsymbol{\xi}) \,\hat{\bo{b}}\, \hat{S}(\boldsymbol{\xi}) &= \cosh(\bo{R})\hat{\bo{b}} - \sinh(\bo{R})e^{i \bo{\Theta}}\hat{\bo{a}}^{\dag}\\
\hat{S}^{\dag}(\boldsymbol{\xi}) \,\dagop{\bo{a}}\, \hat{S}(\boldsymbol{\xi}) &= \cosh(\twid{\bo{R}})\dagop{\bo{a}}-\sinh(\twid{\bo{R}})e^{-i \twid{\bo{\Theta}}}\hat{\bo{b}}\\
\hat{S}^{\dag}(\boldsymbol{\xi}) \,\dagop{\bo{b}}\, \hat{S}(\boldsymbol{\xi}) &= \cosh(\twid{\bo{R}})\dagop{\bo{b}}-\sinh(\twid{\bo{R}})e^{-i \twid{\bo{\Theta}}}\hat{\bo{a}},
	\end{split}
\end{equation}
and their transposes
\begin{equation}\label{eq:BOG_T2}
	\begin{split}
\hat{S}^{\dag}(\boldsymbol{\xi}) \,\twid{\hat{\bo{a}}}\, \hat{S}(\boldsymbol{\xi}) &= \twid{\hat{\bo{a}}}\cosh(\twid{\bo{R}})-\twid{\dagop{\bo{b}}} e^{i \twid{\bo{\Theta}}} \sinh(\twid{\bo{R}})\\
\hat{S}^{\dag}(\boldsymbol{\xi}) \,\twid{\hat{\bo{b}}}\, \hat{S}(\boldsymbol{\xi}) &= \twid{\hat{\bo{b}}}\cosh(\twid{\bo{R}})-\twid{\dagop{\bo{a}}} e^{i \twid{\bo{\Theta}}} \sinh(\twid{\bo{R}})\\
\hat{S}^{\dag}(\boldsymbol{\xi}) \,\twid{\dagop{\bo{a}}}\, \hat{S}(\boldsymbol{\xi}) &= \twid{\dagop{\bo{a}}}\cosh(\bo{R})-\twid{\hat{\bo{b}}} e^{-i \bo{\Theta}} \sinh(\bo{R})\\
\hat{S}^{\dag}(\boldsymbol{\xi}) \,\twid{\dagop{\bo{b}}}\, \hat{S}(\boldsymbol{\xi}) &= \twid{\dagop{\bo{b}}}\cosh(\bo{R})-\twid{\hat{\bo{a}}} e^{-i \bo{\Theta}} \sinh(\bo{R}).
	\end{split}
\end{equation}
With these transformations at our disposal, we can calculate the expectation values of many interesting quantities.
\subsection{Quadrature Variance}
First, we investigate the variances of the quadrature operators, which we define as
\begin{equation}
\begin{split}\label{eq:QUADS}
\hat{\bo{X}}_1 =& \dfrac{1}{2^{3/2}}(\hat{\bo{a}}+\hat{\bo{a}}^{\dagger} + \hat{\bo{b}}+\hat{\bo{b}}^{\dagger})\\
\hat{\bo{X}}_2 =& \dfrac{1}{i 2^{3/2}}(\hat{\bo{a}}-\hat{\bo{a}}^{\dagger} + \hat{\bo{b}}-\hat{\bo{b}}^{\dagger}).
\end{split}
\end{equation}
\subsubsection{Scalar Variance}
We first investigate the scalar variance of the transverse multispatial mode field.
The scalar quadrature variance is defined as
\begin{equation}
\begin{split}
\expt{(\Delta \hat{X}_j)^2} \equiv \expt{\Delta \twid{\hat{\bo{X}}_j} \; \Delta \hat{\bo{X}}_j },
\end{split}
\end{equation}
where $\Delta \hat{\bo{X}}_j \equiv \hat{\bo{X}}_j - \expt{\hat{\bo{X}}_j}$.
In the spontaneous nonlinear interaction regime, that is, vacuum in the signal and idler input modes, we are assured that $\expt{\hat{\bo{X}}_j} = \bo{0}$ and we thus find
\begin{equation}\label{eq:SCALAR_VAR}
\begin{split}
\expt{(\Delta \hat{X}_j)^2} &= \expt{\twid{\hat{\bo{X}}_j} \; \hat{\bo{X}}_j }\\
&=\bra{\{0\}_B,\{0\}_A}\dagop{S} \twid{\hat{\bo{X}}_j} \hat{S} \, \\
&\times \dagop{S} \hat{\bo{X}}_j \hat{S}\ket{\{0\}_A,\{0\}_B},
\end{split}
\end{equation} 
where $\ket{\{0\}_A,\{0\}_B}$ is the vacuum state for the multimode twin beam, and $\ket{\{ 0 \}}=\ket{0_1,0_2,...}$ is the typical multimode vacuum state.
Using Eqs.~(\ref{eq:BOG_T})--(\ref{eq:QUADS}) in Eq.~(\ref{eq:SCALAR_VAR}) we find
\begin{equation}\label{eq:SCALAR_VAR2}
\begin{split}
\expt{(\Delta \hat{X}_{1,2})^2} &= \dfrac{1}{4}\mathrm{Tr}\big\{ \cosh^{2}(\twid{\bo{R}})+\sinh^{2}(\bo{R})\\ 
&\mp \big( \cosh(\twid{\bo{R}}) \sinh(\bo{R}) e^{i \bo{\Theta}} + \sinh(\bo{R}) \cosh(\twid{\bo{R}}) e^{-i \bo{\Theta}} \big) \big\}\\
&= \dfrac{1}{4}\mathrm{Tr}\big\{ \cosh^{2}(\bo{R})+\sinh^{2}(\bo{R})\\
&\mp 2 \sinh (\bo{R}) \cosh (\bo{R}) \cos (\bo{\Theta}) \big\} \\
&= \dfrac{1}{4} \mathrm{Tr} \big\{ \cosh(2\bo{R}) \mp \sinh(2 \bo{R}) \cos (\bo{\Theta}) \big\},
\end{split}
\end{equation}
where in the second line we use the cyclicity of trace, the invariance under transpose, and the commutation properties of $\bo{R}$ and $\bo{\Theta}$ to find a familiar form, and in the final line present a compact form. 
Therefore, we see that the total quadrature variance is simply the sum of all the quadrature variance in each of the transverse multispatial modes. 
\subsubsection{Variance Matrices}
Next, we investigate the corresponding variance-covariance matrices, which can be written as
\begin{equation}\label{eq:MATRIX_VAR}
\begin{split}
\expt{(\Delta \hat{\bo{X}}_j)^2} \equiv \expt{\Delta \hat{\bo{X}}_j \, \Delta \twid{\hat{\bo{X}}_j}} = \expt{ \hat{\bo{X}}_j \, \twid{\hat{\bo{X}}_j} },
\end{split}
\end{equation}
where again we have assumed $\expt{\hat{\bo{X}}_j}=0$.
Using Eqs.~(\ref{eq:BOG_T})--(\ref{eq:QUADS}) in Eq.~(\ref{eq:MATRIX_VAR}) we find
\begin{equation}\label{eq:MATRIX_VAR2}
\begin{split}
\expt{(\Delta \hat{\bo{X}}_{1,2})^2}  &=  \dfrac{1}{8} \big[ \cosh(2\bo{R})+\cosh(2\twid{\bo{R}})  \\
&\mp \big( \sinh(2\bo{R})e^{i \bo{\Theta}} +  \sinh(2\twid{\bo{R}}) e^{-i \twid{\bo{\Theta}}} \big) \big].
\end{split}
\end{equation}
To investigate the correlations between $\hat{\bo{X}}_1$ and $\hat{\bo{X}}_2$, we calculate the cross-covariance matrix 
\begin{equation}\label{eq:COV}
\begin{split}
\mathrm{\textbf{cov}}(\hat{\bo{X}}_1,\,\hat{\bo{X}}_2) &\equiv \dfrac{1}{2}\big( \expt{ \hat{\bo{X}}_1 \, \twid{\hat{\bo{X}}_2} }+\expt{ \hat{\bo{X}}_2 \, \twid{\hat{\bo{X}}_1} }^{T} \big)\\
&= \dfrac{i}{4} \big[ \cosh{2\bo{R}} - \cosh{2 \twid{\bo{R}} }\\
&+ \sinh(2\bo{R}) e^{i \bo{\Theta}}-\sinh(2\twid{\bo{R}}) e^{-i \twid{\bo{\Theta}}} \big],
\end{split}
\end{equation}
which happens to equalize the general uncertainty relation
\begin{equation}\label{eq:UNC_RE}
\expt{(\Delta \hat{\bo{X}}_{1})^2} \, \expt{(\Delta \hat{\bo{X}}_{2})^2} \geq \dfrac{1}{4}\big(\mathrm{\textbf{cov}}(\hat{\bo{X}}_1,\,\hat{\bo{X}}_2)\big)^2 + \dfrac{\bo{I}}{16}.
\end{equation}
When the squeeze matrix is symmetric \textit{and} Hermitian, it has real entries.
Therefore, $\bo{\xi} = \twid{\bo{\xi}}$ and $\bo{\xi}=\bo{\xi}^\dagger \Longrightarrow \bo{R}=\twid{\bo{R}}$ and $e^{i \bo{\Theta}} = e^{-i \twid{\bo{\Theta}}}$. This reduces the variances to 
\begin{equation}
\begin{split}
&\mathrm{cov}(\hat{\bo{X}}_1,\,\hat{\bo{X}}_2)=0\\
&\expt{(\Delta \hat{\bo{X}}_{1})^2} \, \expt{(\Delta \hat{\bo{X}}_{2})^2} = \dfrac{I}{16}.
\end{split}
\end{equation}
Furthermore, when $\bo{\xi}$ is positive semidefinite we have $e^{i \bo{\Theta}}=I$, and Eq.~(\ref{eq:MATRIX_VAR2}) reduces to the simple form 
\begin{equation}
\expt{(\Delta \hat{\bo{X}}_{1,2})^2} =\dfrac{1}{4}e^{\mp 2 \bo{R}}.
\end{equation}
\subsection{Photons and Multimode Correlations}
Now we will investigate the average photon number of the squeezed state $\ket{\xi}$ and the correlations. 
We find
\begin{equation}
\begin{split}
\expt{\hat{n}_a} &= \bra{\xi}\twid{\hat{\bo{a}}^{\dagger}} \hat{\bo{a}} \ket{\xi}=\tr{\sinh^{2}(\bo{R})}\\
\expt{\hat{n}_b} &= \bra{\xi}\twid{\hat{\bo{b}}^{\dagger}} \hat{\bo{b}} \ket{\xi}=\tr{\sinh^{2}(\bo{R})}\\
\expt{\hat{n}_a^2} &= \bra{\xi} (\twid{\hat{\bo{a}}^{\dagger}} \hat{\bo{a}})^2 \ket{\xi} = \dfrac{1}{4} \tr{\sinh^{2}(2\bo{R})} + \tr{\sinh^{2}(\bo{R})}^2 \\
\expt{\hat{n}_b^2} &= \bra{\xi} (\twid{\hat{\bo{b}}^{\dagger}} \hat{\bo{b}})^2 \ket{\xi} = \dfrac{1}{4} \tr{\sinh^{2}(2\bo{R})}  + \tr{\sinh^{2}(\bo{R})}^2,
\end{split}
\end{equation}
which gives the number variance
\begin{equation}
\expt{(\Delta \hat{n}_{a} )^2} = \expt{(\Delta \hat{n}_{b} )^2} = \dfrac{1}{4} \tr{\sinh^{2}(2\bo{R})},  
\end{equation}
each of which reduces to the familiar result for the case of single transverse spatial modes.
In similar fashion, the covariance is 
\begin{equation}
\mathrm{cov}(\hat{n}_a , \hat{n}_b) = \dfrac{1}{4} \tr{\sinh^{2}(2\bo{R})}.
\end{equation} 
Next, we wish to investigate the interspatial-mode photon number correlations in a way that the covariance cannot.
Typically, one would investigate the probability $P_{\ell,p}$ of finding a photon in the $\ell,p$ mode.
However, this information is naturally contained along the diagonal of the average photon number matrix
\begin{equation}
\bar{\bo{n}} = \expt{\hat{\bo{a}}^{\dagger} \twid{\hat{\bo{a}}}} = \sinh^2 (\bo{R}).
\end{equation}
Thus, we calculate the photon-pair creation matrix which reveals the coupling strength between transverse spatial modes of the spatial modes $\hat{\bo{a}}$ and $\hat{\bo{b}}$: 
\begin{equation}\label{eq:PCM}
\bo{M}_{a \leftrightarrow b} \equiv \expt{
\hat{ \bo{a}}^{\dagger} \twid{\hat{\bo{b}}^{\dagger}}}  = \dfrac{1}{2} e^{-i \bo{\Theta}} \sinh(2\bo{R}).
\end{equation}
When normalized, the modulus of the matrix elements give the probability of transverse-spatial modes pairing in the nonlinear interaction, thus containing a photon pair.
\subsection{Eigenmodes of Squeezing}\label{Eigenmodes}
In general, the squeezing matrix is neither symmetric nor Hermitian.
However, under certain conditions, for example, when the beam focal points are at the center of the non linearity, it can be normal.
Therefore, the following analysis is valid, or a good approximation, for many experimental configurations.   
When $\bo{\xi}$ is normal, it can be diagonalized by a unitary.
If we let $\boldsymbol{U}$ be the matrix whose columns are eigenvectors of $\bo{\xi}$, then we can diagonalize $\boldsymbol{\xi}$ according to $\bo{\xi}^{\prime} \equiv \bo{U}^{\dagger}\bo{\xi}\bo{U}$.
Furthermore, the decomposition yields the diagonal matrices $\bo{R}^{\prime}$ and $\bo{\Theta}^{\prime}$.   
The corresponding eigenmodes of squeezing are found according to $\hat{\bo{a}}^{\prime} \equiv \bo{U}^{\dagger} \hat{\bo{a}}$, $\hat{\bo{b}}^{\prime} \equiv \twid{\bo{U}} \hat{\bo{b}}$, $\hat{\bo{a}}^{\dagger \prime} \equiv \twid{\bo{U}} \hat{\bo{a}}^{\dagger}$, and $\hat{\bo{b}}^{\dagger \prime} \equiv \bo{U}^{\dagger} \hat{\bo{b}}^{\dagger}$. 
It follows that their Bogoliubov transformations have a particularly simple form. 
The $i^{\mathrm{th}}$ modes in the eigenmode vectors become 
\begin{equation}\label{eq:BOG_T3}
\begin{split}
\dagop{S} (\bo{\xi}) \, \hat{a}_{i}^{\prime} \, \hat{S}(\bo{\xi}) &= 
\cosh R^{\prime}_i \, \hat{a}_{i}^{\prime} + 
\sinh R^{\prime}_i \, e^{i\Theta^{\prime}_{i}} \, \hat{b}^{\dagger \prime}_{i} \\
\dagop{S} (\bo{\xi}) \, \hat{b}^{\prime}_{i} \, \hat{S}(\bo{\xi}) &= 
\cosh R^{\prime}_i \, \hat{b}^{\prime}_{i} + 
\sinh R^{\prime}_i \, e^{i\Theta^{\prime}_{i}} \, \hat{a}^{\dagger \prime}_{i} \\
\dagop{S}(\bo{\xi}) \, \hat{a}^{\dagger \prime}_{i}\, \hat{S}(\bo{\xi}) &= 
\cosh R^{\prime}_i \, \hat{a}^{\dagger \prime}_{i} +
\sinh R^{\prime}_i \, e^{-i\Theta^{\prime}_{i}}  \, \hat{b}^{\prime}_{i} \\
\dagop{S} (\bo{\xi}) \, \hat{b}^{\dagger \prime}_{i} \, \hat{S}(\bo{\xi}) &= 
\cosh R^{\prime}_i \, \hat{b}^{\dagger \prime}_{i} +
\sinh R^{\prime}_i \, e^{-i\Theta^{\prime}_{i}} \, \hat{a}^{\prime}_{i} ,
\end{split}
\end{equation}
where $R^{\prime}_i$ are the diagonal elements of $\bo{R}^{\prime}$, and $\Theta^{\prime}_{i}$ are the diagonal elements of $\bo{\Theta}^{\prime}$.
Thus, the eigenmodes of squeezing are fundamental in the sense that they transform according to the canonical two-mode squeezed-vacuum equations~\cite{bennink2002improved}.
They also define a basis in which to analyze the squeezing and determine which modes are squeezed the most.
The largest $\lambda_i$ corresponds to the largest multimode squeezing and $\hat{a}^{\prime}_{i}$ gives that collection of modes.

The question remains as to whether we can use the eigenmodes of squeezing as a basis to represent our squeezed state $\ket{\bo{\xi}}$. 
Thus, we first would like to test whether the eigenmodes satisfy the canonical commutation relation.
Evidently,
\begin{equation}
[\hat{a}^{\prime}_{i},\hat{a}^{\dagger \prime}_{j}] = U^{\dagger}_{ik} U_{kj} = \delta_{ij}, 
\end{equation}
and we find that the squeezed state takes the particularly simple form
\begin{equation}
\ket{\bo{\xi}}_{\lambda} = \sum_{i,n} \mathrm{sech}(\lambda_i) \tanh^n (\lambda_i) \ket{\{ n\}_i}_A \ket{\{ n\}_i}_B,
\end{equation}
where $A,B$ indicate the two spatial modes $\hat{\bo{a}},\hat{\bo{b}}$, respectively, and $\ket{\{ n\}_i}$ is the multimode Fock state with $n$ photons in the $i^{\mathrm{th}}$ eigenmode. 
We use these states as our basis states since $\expt{\dagop{a}_\lambda \dagop{b}_{\lambda^{\prime}} } = \delta_{\lambda, \lambda^{\prime}}$, in other words, the photons are created pairwise in the same eigenmodes. 

\section{Simulations}
\begin{figure}[!b]
	\includegraphics[width=\columnwidth]{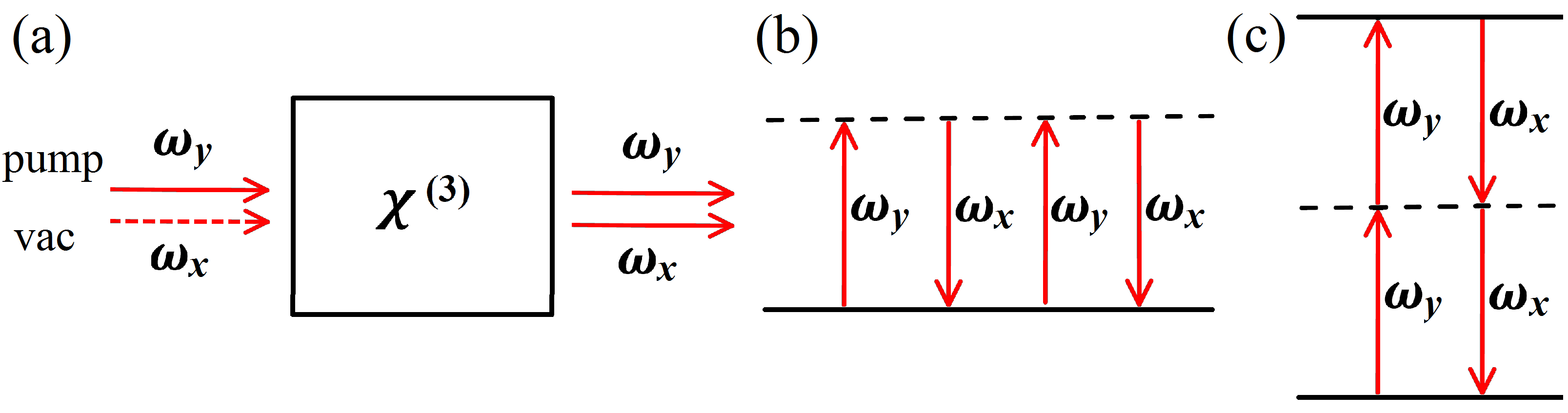}
	\caption{\label{fig:PSR_Schematic} Setup and energy diagrams describing the PSR effect in $\mathrm{^{87}Rb}$. In (a) we show a linear polarized pump beam interacting with vacuum fluctuations in the orthogonal polarization. Two tensor elements of the semiclassical susceptibility tensor survive: (b) depicts the single-photon resonance associated with the $\chi_{xyxy}$ element, that is, a response in $x$ due to stimulation in $yxy$, and (c) describes the two-photon resonance associated with the $\chi_{xyyx}$ element, that is, a response in $x$ due to stimulation in $yyx$. The input and output fields are co-propagating.   
	}
\end{figure}
The preceding theory is a powerful tool can be used to understand the complex TSM structure stimulated during nonlinear-optical interactions. In fact, some interactions may not require the full capability of our theory. To that end, we simulate several interactions, progressing from simple to complex, demonstrating the utility of this theory as an analysis tool.     
First, we investigate the polarization self-rotation effect observed in $\mathrm{^{87}Rb}$ (see Fig.~\ref{fig:PSR_Schematic}).
The tensor nature of this $\chi^{(3)}$ interaction is the fundamental phenomenon related to this effect, and it is worked out in detail for classical fields \citep{Boyd_book}.
Quantized treatments have also been performed which relate the observed noise suppression to the amount of polarization self-rotation observed in the medium \cite{matsko2002vacuum}.
The early work predicted levels of noise suppression which have proven to be woefully over optimistic.
This realization prompted more rigorous noise calculations \cite{lezama2008numerical} and our experimental study of the transverse spatial modes excited during the interaction \cite{zhang2016spatial, zhang2017multipass}.
With our preceding theory, we are finally able to perform a fully second-quantized analysis of the transverse-spatial-mode structure.
For brevity, and to keep the proceeding relatively straightforward, we will analyze the PSR effect in terms of the resonance structure of the interaction.
In other words, we will not consider the spatial structure of $\chi$.   

The classical formulation for general third-order interactions divides the interaction into single- and two-photon resonant contributions, each with different photon polarization interaction processes (see Fig.~\ref{fig:PSR_Schematic}).
Thus, the following simulations are separated into single- and two-photon resonant interactions, and serve as a stepping stone to investigating other four-wave mixing and down-conversion processes.     

\begin{figure}[!b]
	\includegraphics[width=\columnwidth]{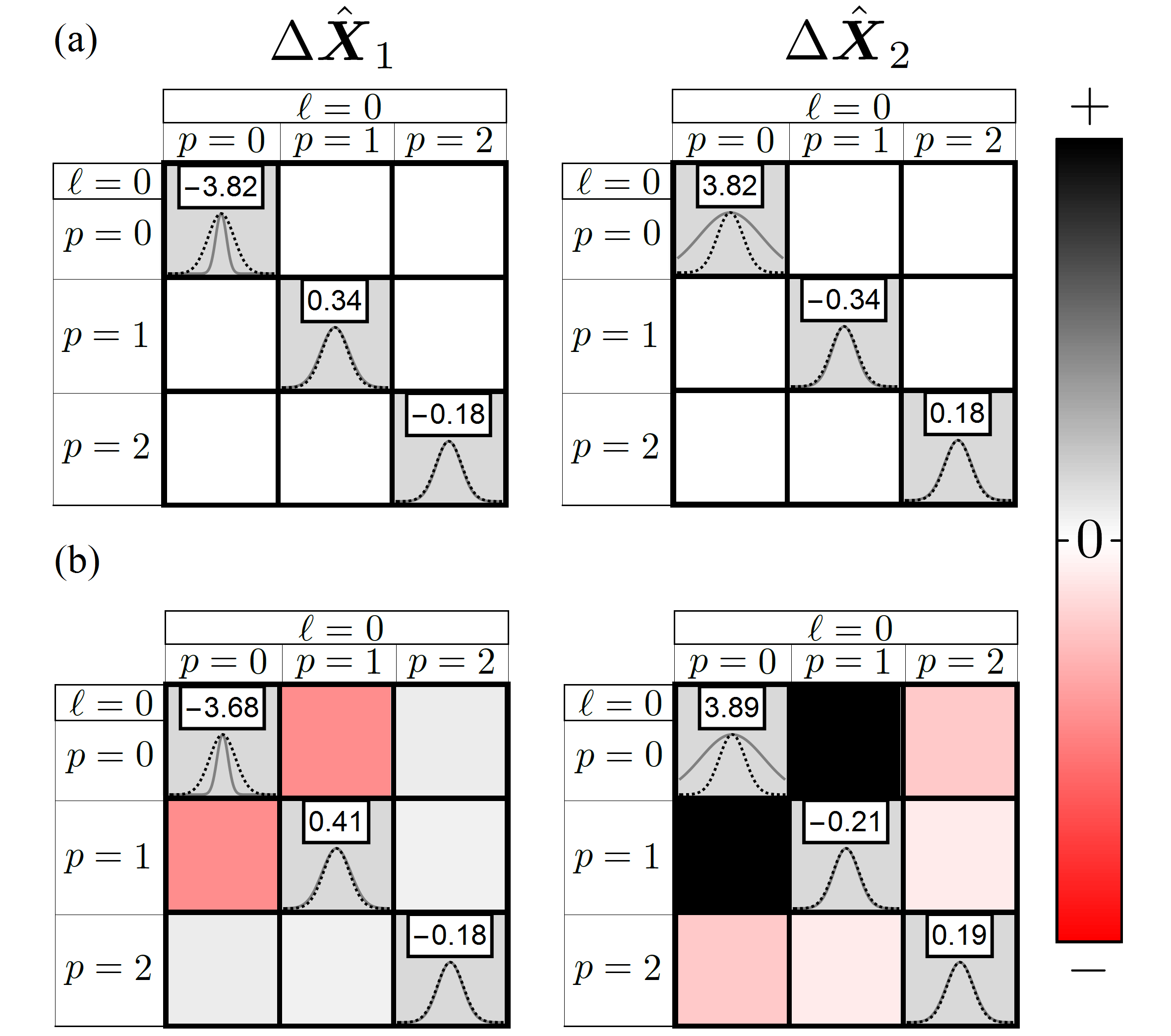}
	\caption{\label{fig:PSR1_2} Noise matrices for the single-spatial mode quadratures $\hat{\bo{X}}_1$ and $\hat{\bo{X}}_2$, in (a) the absence of TSM cross talk and (b) $p$-mode crosstalk . The noise in the quadrature is depicted along the diagonal as the projection of the multimode Wigner function onto the particular TSM quadrature. The dashed black Gaussian is the projection of the vacuum state, thus the diagonal elements allow one to quickly observe how the noise suppression, if at all, is distributed among the spatial modes. The inset number is the amount of squeezing given in decibels. The off-diagonal elements represent the covariance between different spatial modes in the quadrature. In (a) there is no cross talk between the x-polarization modes, thus the covariance is zero. In (b) the $p$-mode cross talk is present and indicated by the nonzero off-diagonal elements. 
	}
\end{figure} 
The first and simplest case to consider is the single-photon resonance scheme in Fig.~\ref{fig:PSR_Schematic}(b).
Since the photons scattered into the $x$ polarization are separated by $y$-polarization excitation, it is reasonable to assume that the two $x$-polarization photons do not have transverse-spatial-mode correlations, that is, there is no cross talk between the two-photon emissions.
Mathematically, we take Eq.~(\ref{eq:SQUEEZER}), let $\hat{\bo{b}} \rightarrow \hat{\bo{a}}$, and thus have a single-spatial-mode squeezer.
We use this calculation as a baseline simulation with which to compare the more sophisticated interactions. 
Thus, we scale the strength of the interaction such that $\bar{n} = 1$, and examine how $\bar{n}$, $\Delta \hat{\bo{X}}_1$, $\Delta \hat{\bo{X}}_2$, and $\bo{M}_{a \leftrightarrow b}$ change in each case. 

\subsection{Four-Wave Mixing}
\begin{figure*}[!t]
	\includegraphics[width=2\columnwidth]{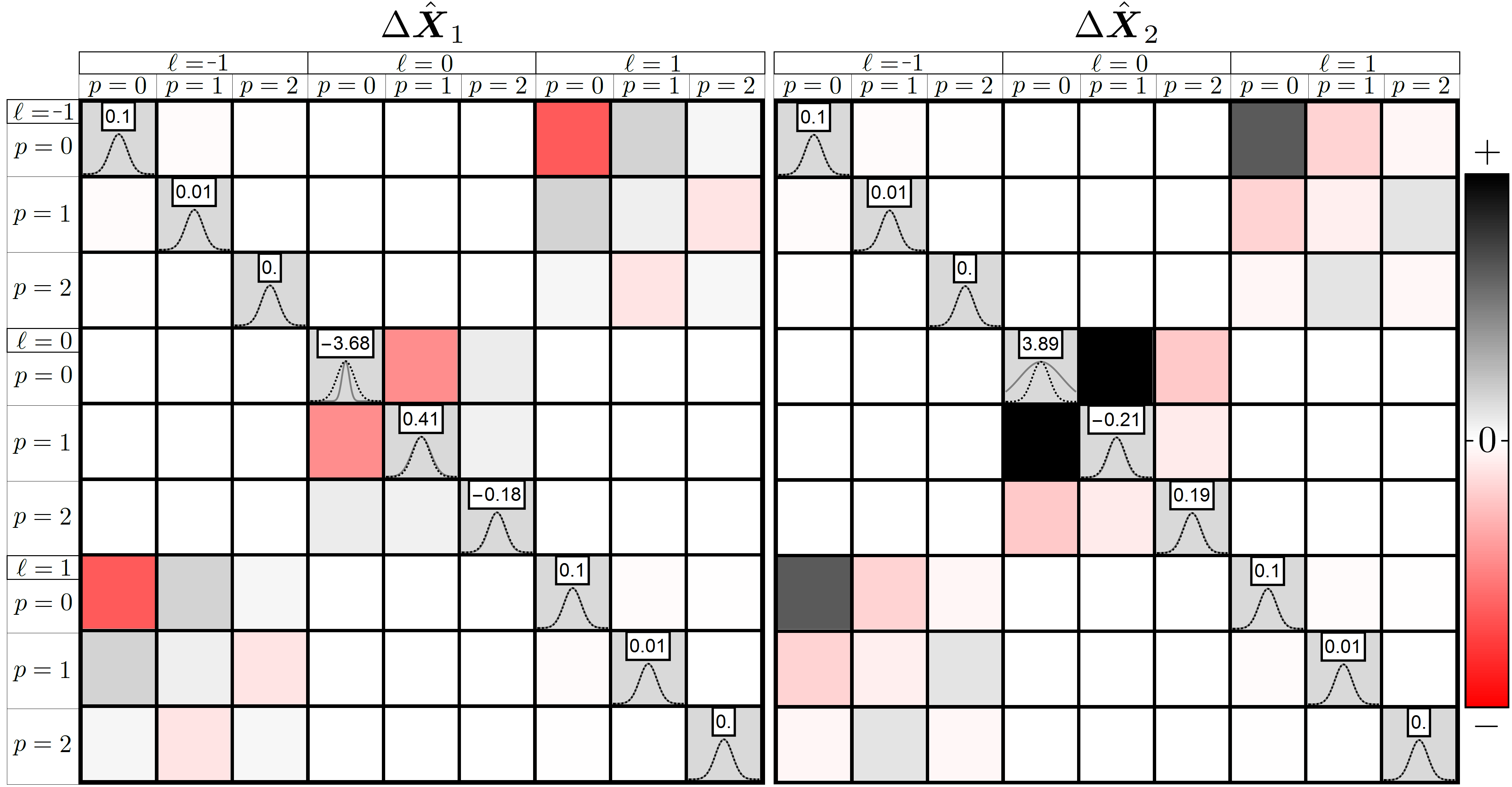}
	\caption{\label{fig:PSR3} Noise matrices for the joint quadratures $\hat{\bo{X}}_1$ and $\hat{\bo{X}}_2$, when full cross talk is present in a two-photon resonance scheme. The noise in the quadrature is depicted by the solid Gaussian along the diagonal, and the off-diagonal elements represent the covariance between different spatial modes in the quadrature. Red is negative, white is zero, and black is positive covariance. The inset gives the squeezing in decibels. 
	}
\end{figure*}
For the first simulation, we assume a 795-nm pump beam, in a $u_{0,0}$ mode with a 80-$\mu$m waist, is focused at the center of a 3$z_R$ long nonlinear cell.
First, we will examine the quadrature noise matrices [see Fig.~\ref{fig:PSR1_2}(a)].
As a visualization tool, the plots along the diagonal represent the quadrature noise.
The gray Gaussian represents the projection of the multimode Wigner function onto the particular TSM quadrature.
The dashed black Gaussian is the projection of the vacuum Wigner function and serves as a reference.
Thus we can quickly observe how the noise suppression, if at all, is distributed among the spatial modes, and the inset number gives the amount of squeezing given in decibels.
The off-diagonal elements of the quadrature matrices are the covariance, and the strength of covariance is given by the color map in the bar legend.
As expected, there is no response in $\ell\neq0$ modes and we see a slight amount of squeezing in the higher-order $p$ modes, but it is mostly concentrated in the $u_{0,0}$ mode.
Several of our studies, using spatial masks \cite{zhang2016spatial} and optimization procedures \cite{zhang2017multipass}, reveal the absence of any azimuthal structure (that is, $\ell=0$) and subtle pollution from higher-order $p$ modes.  
But what if cross talk between the $p$ modes is taking place?
There would still be no azimuthal structure to detect, but there would indeed be a more complex $p$-mode structure that will effect the squeezing in the system.

\subsubsection{PSR with Single-Photon Resonance}
Next, we allow for cross talk between $p$ modes in the $x$ polarization.
Mathematically, this means we now accommodate for the two spatial modes $\hat{\bo{a}}$ and $\hat{\bo{b}}$, as in Eq.~(\ref{eq:SQUEEZER}), and insert the restriction $\delta_{\ell,m}$.   
In Fig.~\ref{fig:PSR1_2}(b) we show the quadrature noise matrices for this simulation.
Although subtle in the figure, there is actually $\small\sim 3 \%$ increase in quadrature noise in the $u_{0,0}$ squeezed mode and $\bar{n} \sim 1.14$.
This trend agrees with previous findings , that in general the population of higher-order modes will deteriorate the performance of non classical processes, including squeezing \cite{zhang2016spatial}.
Furthermore, we now observe covariance between TSMs, the color indicating whether the variances are positively or inversely related.
Thus, we observe the amount of correlations between the modes in each quadrature, which is a prelude to examining the coupling strength quantified by the coupling matrix $\bo{M}_{a \leftrightarrow b}$.
This we save for the next section, where we show that the mode structure is in general much more complicated, by simulating a generic four-wave mixing scheme with a two-photon resonance. 

\subsubsection{FWM with Two-Photon Resonance}
The two-photon resonance in Fig.~\ref{fig:PSR_Schematic}(c) will lead to a more complicated mode structure, since the two $x$-polarized photons are emitted in cascade, allowing full cross talk.
For example, it is well known that even when the pump beam carries no orbital angular momentum (OAM), the scattered photons can in principle carry opposite $\ell$, thus conserving OAM. 
This, of course, happens much less frequently than excitation in the $u_{0,0}$ mode, which dominates because of the ideal overlap with the pump~\cite{lanning2017gaussian}.
Our theory allows us to investigate this complicated mode structure. 
As expected, this simulation shows a response at $\pm \ell$ along with the $p$ mode structure (see Fig.~\ref{fig:PSR3}).
Surprisingly, there is no increase in noise of the $u_{0,0}$ mode, despite the more complicated structure and the average photon number increasing to $\bar{n}=1.25$.
Another qualitative observation is the obvious correlation between positive covariance with the noisy quadrature and negative covariance with the squeezed quadrature.   
It seems that overall cooperative noise fluctuations in quadrature correspond directly to excess noise, whereas opposing noise fluctuations among the spatial corresponds to noise suppression.

\begin{figure}[!t]
	\includegraphics[width=1.0\columnwidth]{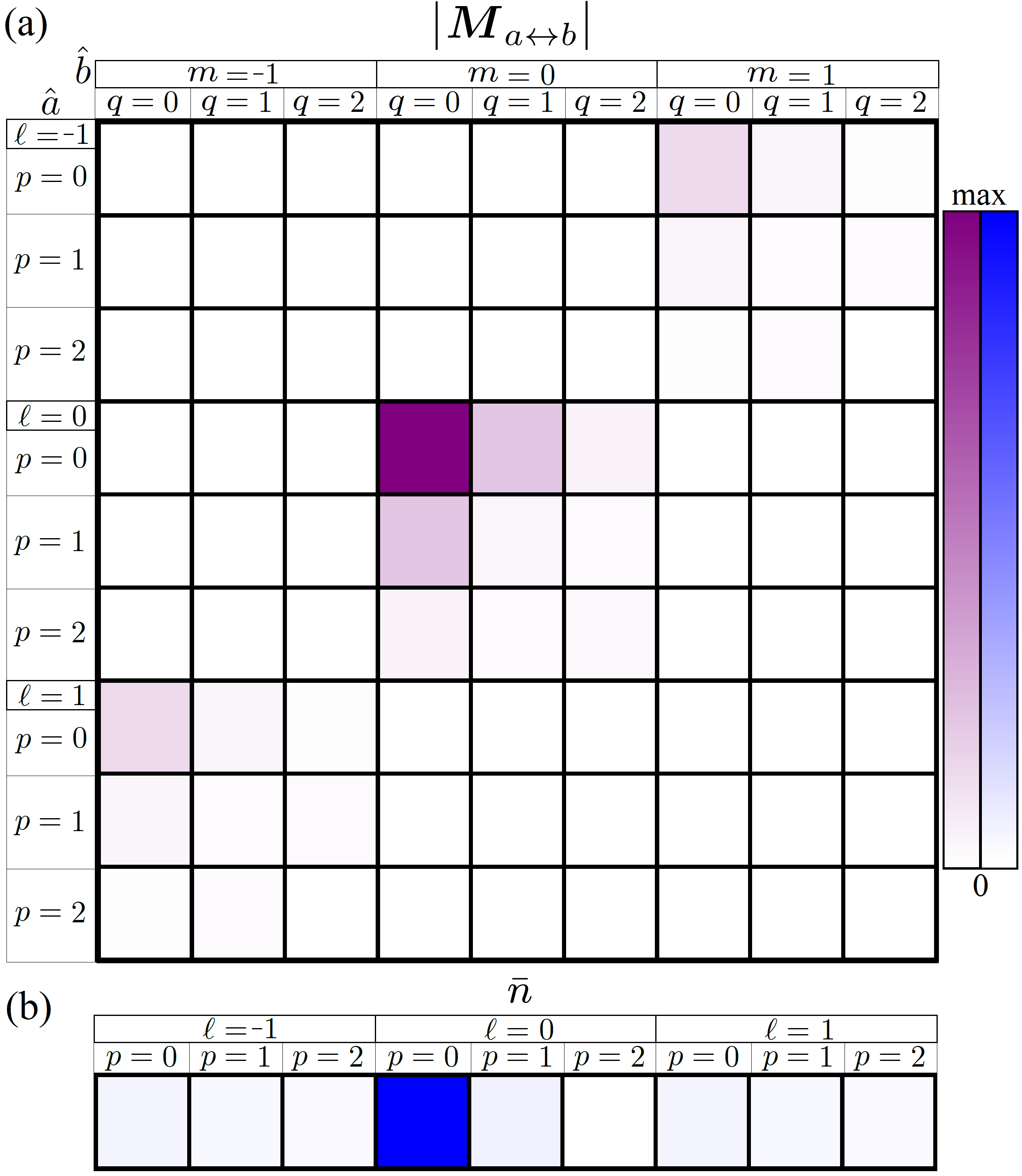}
	\caption{\label{fig:PSR3_M_n} (a) Photon-creation matrix given by Eq.~(\ref{eq:PCM}) and (b) average photon number per mode for degenerate four-wave mixing in a two-photon-resonant scheme where full TSM cross talk is present. For the blue scale, max$=1.25$.     
	}
\end{figure}
In Fig.~\ref{fig:PSR3_M_n}(a) we plot a histogram of the photon-creation matrix $\bo{M}_{a \leftrightarrow b}$ given by Eq.~(\ref{eq:PCM}).
It shows the mode structure of the quantum beam and identifies which \textit{pairs of modes} are most likely to be populated.
Response along the diagonal indicates that the photon pairs are created in identical TSMs.
Contrastingly, off-diagonal response indicates that photon pairs can be excited in different TSMs.  
In Fig.~\ref{fig:PSR3_M_n}(b) we plot a histogram of the diagonal elements of $\expt{\bo{n}} \equiv \expt{ \dagop{\bo{a}} \, \twid{\hat{\bo{a}}} }$, that is, $\bar{n}$ per mode.
This is a relatively simple system and the parameters of the interaction have already been optimized for concentrating squeezing in a single mode.
In general, there will be much more cross talk and this theory can be used as a tool to tailor the quantum-mode structure.   
Furthermore, this type of analysis is a very convenient tool for systems heralding a single photon, or creating an indistinguishable entangled photon pair. 
However, knowing which mode is squeezed the most, or which mode has more photons on average, will not necessarily allow one to maximize the performance of their process.
For this we will simulate a parametric down-conversion process and show how one can recover from the negative effects of the higher-order mode structure.

\subsection{Parametric Down Conversion}
\begin{figure}[!t]
	\includegraphics[width=1.0\columnwidth]{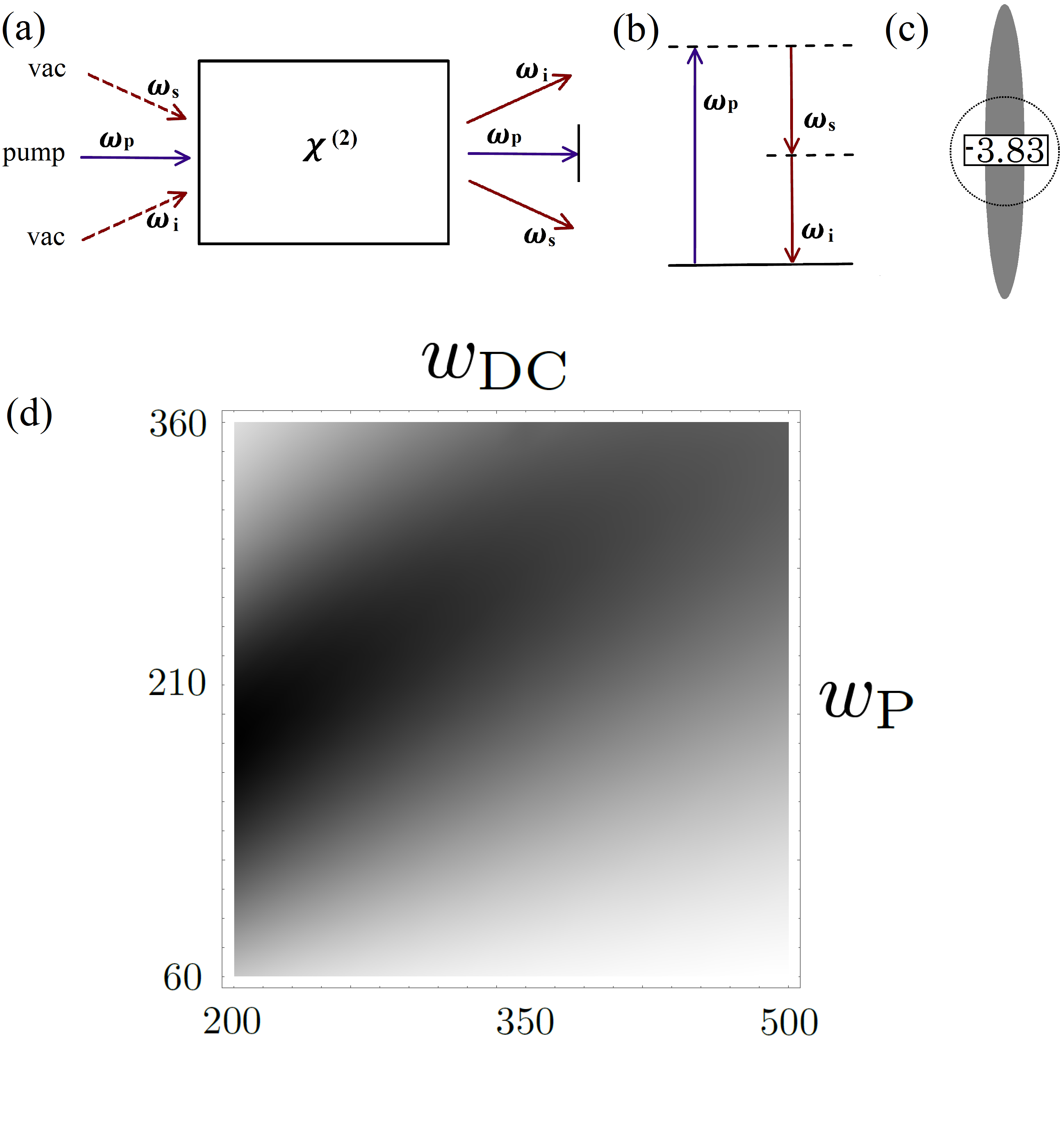}
	\caption{\label{fig:PDC_Schematic_OPT} (a) Schematic of our PDC simulation, (b) the energy-level diagram, (c) the noise ellipse of an ideal squeezed state with $\bar{n}=1$ (inset gives squeezing in decibels), and (d) the density plot, discussed in the main text, which reveals a parameter island where (0,0) coupling is dominant \textit{and} most of the down-converted photons are sent to the $u_{0,0}$ modes. This plot shows $ |M_{00}| / \sum_{i,j}|M_{ij}| \times n_{00} / \sum_{i} n_{ii}$ as a function of the pump waist $w_\mathrm{P}$ and PDC detection waist $w_\mathrm{DC}$ in $\mu$m.   
	}
\end{figure} 
In this section, we simulate type-II down-conversion of 405-nm light into two 810-nm photons [see Figs.~\ref{fig:PDC_Schematic_OPT}(a) and \ref{fig:PDC_Schematic_OPT}(b)].
Ideally, the down-converted photons would share the same transverse spatial mode as the pump photon.
As a visual reference, in Fig.~\ref{fig:PDC_Schematic_OPT}(c) we plot the quadrature noise for an idealized two-mode squeezed-vacuum state with $\bar{n}=1$.
We know, however, that things are not so simple.
In actuality, there is a complex interaction among the transverse spatial modes.

To demonstrate this, we simulate this interaction, allowing full cross talk among the spatial modes but scaling the overall strength of the nonlinearity such that $\bar{n}$ remains 1.
In this way, we can see how the noise suppression may leak into the other transverse spatial modes.
To set a benchmark, we must choose the pump beam waist and the PDC photon collection waist to investigate. 
Suppose then that we want to find the combination of waists that has the strongest response in the $u_{0,0}$ mode.
To do so we calculate $ |M_{00}| / \sum_{i,j}|M_{ij}| \times n_{00} / \sum_{i} n_{ii}$, where $M_{ij}$ and $n_{ij}$ are matrix elements of $\bo{M}_{a \leftrightarrow b}$ and $\bo{n}$ respectively.
Varying the pump and PDC collection waists over a wide experimental range, we find the results, in Fig.~\ref{fig:PDC_Schematic_OPT}(d), which reveal an optimal parameter region.
Thus, we choose $w_{\mathrm{P}}=w_{\mathrm{DC}}=200$-$\mu$m as our benchmark simulation.
In the following, we limit the LG parameters to $-1\leq \ell\leq 1$ and $0\leq p\leq 2$ for display purposes.  

\begin{figure*}[]
	\includegraphics[width=2.0\columnwidth]{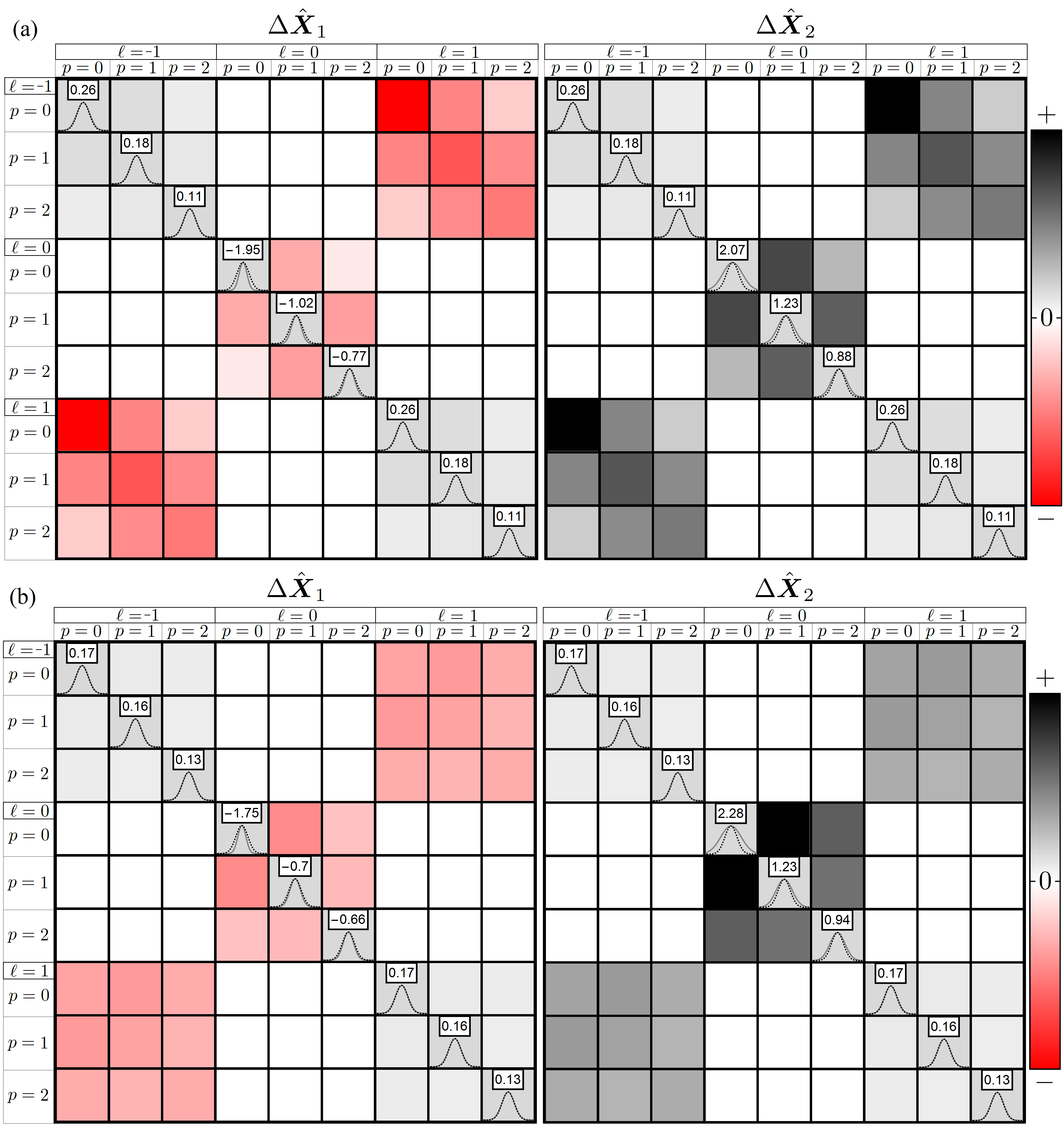}
	\caption{\label{fig:PDC1_2}  (a) Noise matrices for the joint quadratures $\hat{\bo{X}}_1$ and $\hat{\bo{X}}_2$, when full cross talk is present in a type-II PDC scheme, and (b) the noise matrices when the crystal is pumped by the first eigenmode. The noise in the quadrature is depicted by the solid Gaussian along the diagonal, and the off-diagonal elements represent the covariance between different spatial modes in the quadrature. Red is negative, white is zero, and black is positive covariance. In (a), although $\bar{n}=1$, we can see a reduction in the maximum squeezing performance, compared with the ideal situation of only a single transverse spatial mode. In (b) we see that the squeezing changed marginally but the covariance elements experienced the most significant change. This suggests the mode coupling has been altered, and this can be seen clearly by comparing Figs.~\ref{fig:PDC1_M_n_DIAG} and Fig.~\ref{fig:PDC1_M_n_DIAG}(e). We limit $-1\leq \ell \leq 1$ and $0\leq p \leq 2$ for display purposes.     
	}
\end{figure*}

\begin{figure*}[]
	\includegraphics[width=2.0\columnwidth]{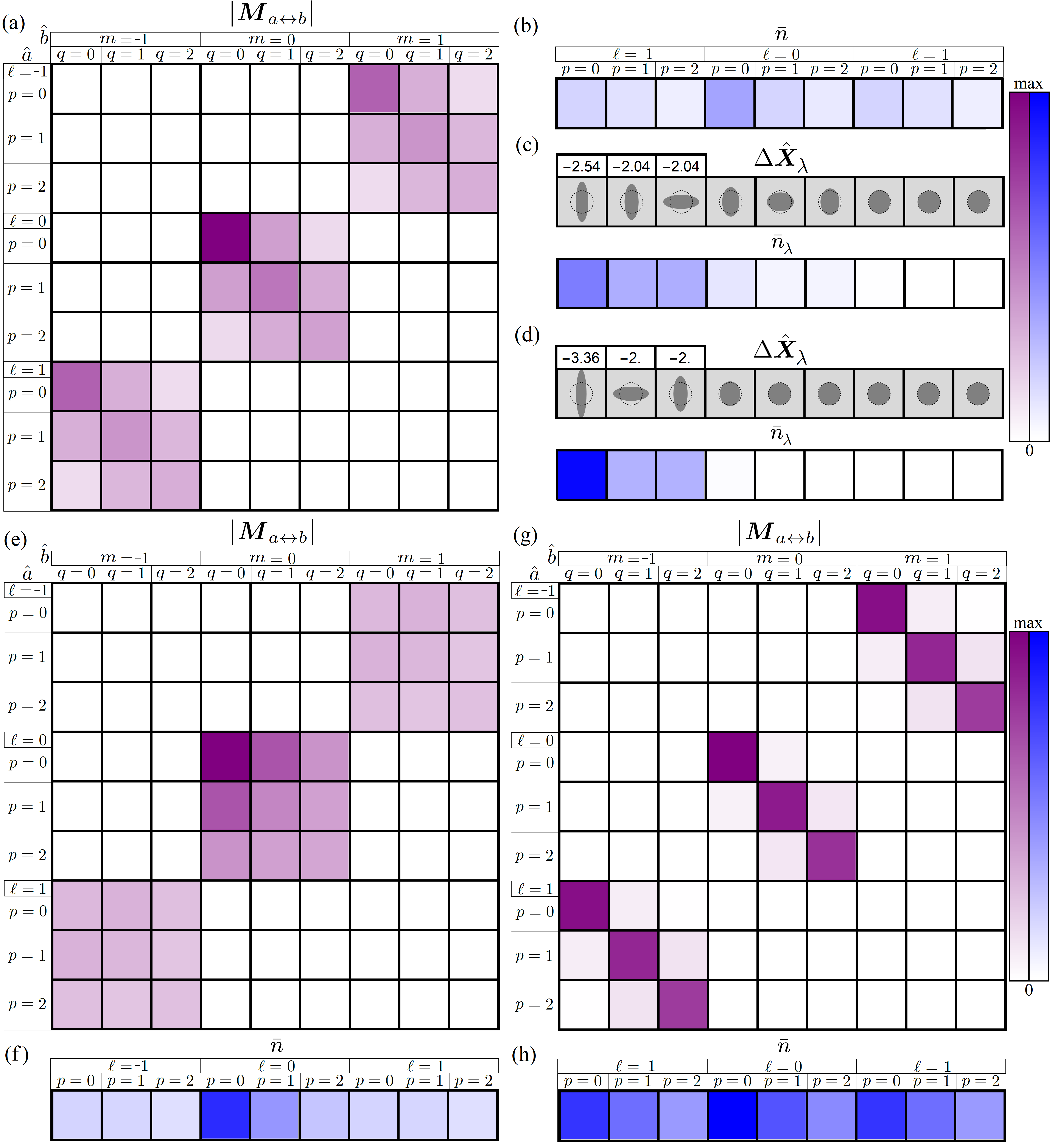}
	\caption{\label{fig:PDC1_M_n_DIAG} (a) Histogram of the photon-creation matrix elements (a.u.) given by Eq.~(\ref{eq:PCM}), (b) the average photon number per mode for our PDC scheme [corresponding to Fig.~\ref{fig:PDC1_2}(a)], and (c, d) the noise ellipses and average photon number for each of the eigenmodes of squeezing, progressing from left to right, starting with $\lambda_1$. In (c) the crystal is pumped with a $u_{0,0}$ mode and in (d) with the $\lambda_1$ eigenmode [corresponding to Fig.~\ref{fig:PDC1_2}(b)]. In (d) we observe an increase in the amount of noise suppression and a shift in the concentration of $\bar{n}_{\lambda}$ toward the first eigenmode. The blue scale is set to max$=0.75$ for (b)-(d). In (e,f) the crystal is pumped with the $\lambda_1$ eigenmode. In (g,h) the pump mode is a $u_{0,0}$ mode but the waist has been increased to 400 $ \mu$m. This configuration is ideal for heralding efficiency since it suppresses the off-diagonal coupling. In other words, it insures that the photon pairs are in the same TSM. The blue scale is set to max$=0.5$ for (b) and max$=0.1$ for (d).
	}
\end{figure*}

In Fig.~\ref{fig:PDC1_2}(a) we show the noise matrices for the joint quadratures $\hat{\bo{X}}_1$ and $\hat{\bo{X}}_2$ using the optimal waists.
One can see that the noise and covariance are qualitatively similar to the two-photon resonant simulation in Fig.~\ref{fig:PSR3}.  
However, the squeezing has leaked into the higher-order modes more drastically.
Furthermore, in Figs.~\ref{fig:PDC1_M_n_DIAG}(a) and \ref{fig:PDC1_M_n_DIAG}(b) we see that 
the cross talk between the modes has also been enhanced, but this is not necessarily a good thing.
For example, in a homodyne measurement, the observed noise suppression will just be the sum of all the noise from each mode that overlaps the local oscillator. 
Thus, leakage into higher-order modes can be detrimental to noise suppression as a resource.  
To recover from this leakage, one can projectively filter out the modes with the most squeezing.
For example, we can see that the $u_{0,0}$ mode has the most squeezing, and thus a naive approach might use a single mode fiber to isolate this mode.
However, better squeezing can be extracted by using the eigenmodes of squeezing.

Employing our theory in Sec.~\ref{Eigenmodes}, we can find the collection of transverse spatial modes which have quadrature noise suppression beyond what is observed in the $u_{0,0}$ mode alone.
This is possible because our choice of quadratures was somewhat arbitrary, in other words, it does not take into account the squeezing parameter and squeezing angle of each mode.
The eigenmode approach, in effect, judiciously chooses the proper quadrature measurement for each transverse spatial mode and thus improves the observed noise suppression. 
This approach may seem mysterious and unenlightening, but the alternative of calculating the Wigner function for this multimode Gaussian squeezed state is quite a difficult problem and an open area of research in itself.
Thankfully, the noise suppression in each eigenmode obeys the canonical two-mode squeezed-vacuum equation; thus we can plot the actual noise ellipse for each of the eigenmodes $\Delta \hat{\bo{X}}_{\lambda}$ along with the average photon number per eigenmode $\bar{n}_{\lambda}$ [see Fig.~\ref{fig:PDC1_M_n_DIAG}(c)]. 
The variance $\Delta \hat{\bo{X}}_{\lambda=1} \sim 0.28$ as compared to $0.32$ in the $u_{0,0}$ mode, an $\sim 0.6$-dB reduction.     
Therefore, one can access an increased amount of noise suppression by detecting the $\lambda_1$ eigenmode.

Next we will investigate how pumping the crystal with an eigenmode effects the noise suppression and mode structure. 
The largest eigenvalue is $\lambda_1$; thus we use this eigenmode for our simulation.
Mathematically, we prepare the pump beam in the superposition $\sum_i [\bo{U}^{\dagger} \bo{u}]_{1i} $, where $\bo{u}$ is a vector of LG modes with the same structure as Eq.~(\ref{eq:VECTORS}).
In Figure~\ref{fig:PDC1_2}(b) we show the noise matrices for the joint quadratures $\hat{\bo{X}}_1$ and $\hat{\bo{X}}_2$. 
The most striking change is the covariance, which has become mostly uniform over all the modes.
Furthermore, we see that there has not been any reduction in noise suppression in any of the LG modes individually. 
However, in Fig.~\ref{fig:PDC1_M_n_DIAG}(d) we see how the tailored pump beam can increase the amount of noise suppression in the first few eigenmodes and shift each $\bar{n}_{\lambda}$ toward the first eigenmode.
Now, the variance is $\Delta \hat{\bo{X}}_{\lambda=1} \sim 0.23$ as compared to $0.32$ in the $u_{0,0}$ mode initially, a $\sim 1.4$~dB reduction.

The change in mode structure is best understood by comparing the photon-creation matrices in Fig.~\ref{fig:PDC1_M_n_DIAG}(a) and \ref{fig:PDC1_M_n_DIAG}(e).
In Figure~\ref{fig:PDC1_M_n_DIAG}(e) see that response in the $\ell = \pm 1$ modes has been suppressed and the coupling in the $\ell=0$ modes has been enhanced.
Furthermore, we see that $\bar{n}$ per mode has shifted toward the $u_{0,0}$ mode.
The same is true for pumping with other eigenmodes, except the shifts may be different. 
For example, pumping with the $\lambda_3$ eigenmode enhances the coupling of the $\ell = \pm 1$ modes and suppresses $\ell = 0$.
This exercise demonstrates that even if the eigenmodes can not be collected, they can suggest ways to change the pump beam to tailor the mode structure of the PDC twin beam.

Next, suppose that we want to maximize the heralding efficiency in a PDC experiment using single-mode fibers.
Then for every (0,0) photon that is detected in the $\hat{\bo{a}}$ mode, we need a matching (0,0) photon in the $\hat{\bo{b}}$ mode.
Now, take the previous PDC simulation as an example.
Examining Fig.~\ref{fig:PDC1_M_n_DIAG}(e), we see that that if a (0,0) photon in the $\hat{\bo{a}}$ mode heralds a photon in the $\hat{\bo{b}}$ mode, there is a significant chance that it will actually be a higher-order $p$ mode.
Therefore, it would be rejected by the single-mode fiber and suppress the heralding efficiency.

So we see that we need to judiciously choose the pump mode structure and collection waist to suppress the off-diagonal elements.      
To do so, we find that we can pump with a (0,0) mode and increase the waist to 400 $\mu$m, keeping the collection waist at 200 $\mu$m. 
As we see in Fig.~\ref{fig:PDC1_M_n_DIAG}(g), doing so suppresses the off-diagonal elements but it is  at the expense of drastically increasing the coupling to higher-order modes.
In fact, the coupling extends far outside the range of this plot, up to $p=20$. 
Furthermore, in in Fig.~\ref{fig:PDC1_M_n_DIAG}(h) we see that the $u_{0,0}$ mode no longer dominates the interaction.
Therefore, concluding our investigation, we see how the heralding efficiency can be enhanced at the expense of the heralding rate.
Although this result is not new, it provides a different theoretical explanation and insight into this body of work.
\section{Conclusion}
We have developed a second quantization procedure which predicts the transverse-spatial-mode structure of quantum beams created in nonlinear optical interactions. 
We used this theory to predict the variance, covariance, and relative coupling strength between the modes. 
Furthermore, we identify the eigenmodes of the interaction and use these to show how they can be used to enhance the noise suppression observed in the system and manipulate the mode coupling.
To utilize the theory, we simulate several interactions, including polarization self-rotation, four-wave mixing, and parametric down-conversion. 
In each case, we concentrate on exposing the underlying transverse-spatial-mode structure, suggesting ways to tailor it by changing the properties of the pump beam, and enhancing the quantum resources by changing the properties of the detection scheme.
The theme of these simulations is enhancing quadrature squeezing or single-photon heralding protocols. 
In these schemes, optimization typically leads to a simplification of the mode structure, since in general, the cross talk of the modes is detrimental to these processes.

The next step is to investigate how to harness the complicated mode structure instead of suppressing it. 
Therefore, we will analyze the properties of OAM path entanglement in several spontaneous nonlinear interactions and suggest how this (potentially higher-order) entanglement can be used as a resource. 
This will be done in two contexts. 
First, we analyze the connectivity of the state, thus determining the utility in cluster-state quantum computing protocols. 
Second, we will calculate the entropy of entanglement, which is a signature of entanglement that gives the number of qubits that can be distilled from the state. 
Once these quantities are known, we will develop new optimization procedures to show how one can improve the performance of these protocols.

This research was supported by the Air Force Office of Scientific Research through Grant No. FA9550-13-1-0098.
R.N.L., Z.X. and J.D.P. would like to acknowledge additional support from the Army Research Office, the Defense Advanced Research Projects Agency, the National Science Foundation, and the Northrop Grumman Corporation.
M.Z., I.N., and E.E.M. acknowledge additional support from the National Science Foundation through Grant No. PHY-308281. 

\bibliography{bibliography/bibliography}

\clearpage
\appendix

\section{Two Photon Amplitude Matrix}
%\subsection{Two Photon Amplitude Matrix}
First, we assume there is some $\ell_\mathrm{max} = m_\mathrm{max}$, and the sums run symmetrically over the azimuthal modes, that is, $-\ell_\mathrm{max} \leq \ell \leq \ell_\mathrm{max} $ and $-m_\mathrm{max} \leq m \leq m_\mathrm{max} $.
Likewise, there is some $p_\mathrm{max} = q_\mathrm{max}$ which determines $0 \leq p \leq p_\mathrm{max}$ and $0 \leq q \leq q_\mathrm{max}$.
We define the two-photon amplitude matrix in such a way that it extends over the azimuthal modes, from negative to positive, and increments the radial index along the way.
One can construct it according to 
\begin{equation}\label{eq:TPA}
\bo{\chi} \equiv \hat{e}_{i+1} \otimes \hat{e}_{j+1} \, \chi_{i - \ell_{\mathrm{max}},j;k-v_{\mathrm{max}},w} \, [\hat{e}_{v+1} \otimes \hat{e}_{w+1}]^{T},
\end{equation}
where $\hat{e}_i$ is a vector with one in the $i$th position, and the sum runs over $0 \leq \i \leq 2 \ell_{\mathrm{max}} $, $0 \leq j \leq p_{\mathrm{max}}$, $0 \leq k \leq 2 m_{\mathrm{max}}$, and $0 \leq w \leq q_{\mathrm{max}}$.
This operational notation may not be clear, so we also include $\bo{\chi}$ in matrix notation:
\begin{widetext}
\begin{equation}
\boldsymbol{\chi}=\begin{pmatrix}
 &\chi_{\ell,p;m,q}\; \; &\chi_{\ell,p;m,q+1}  &\cdots &\chi_{\ell,p;m+1,q}\; \; &\chi_{\ell,p;m+1,q+1} &\cdots \\
 &\chi_{\ell,p+1;m,q}\: \: &\chi_{\ell,p+1;m,q+1} &\cdots &\chi_{\ell,p+1;m+1,q}\; \; &\chi_{\ell,p+1;m+1,q+1} &\cdots \\
&\vdots &\vdots & &\vdots &\vdots &\\
 &\chi_{\ell+1,p;m,q} &\chi_{\ell+1,p;m,q+1} &\cdots &\chi_{\ell+1,p;m+1,q} &\chi_{\ell+1,p;m+1,q+1} &\cdots\\
 &\chi_{\ell+1,p+1;m,q}  &\chi_{\ell+1,p+1;m,q+1} &\cdots  &\chi_{\ell+1,p+1;m+1,q} &\chi_{\ell+1,p+1;m+1,q+1} &\cdots\\
&\vdots &\vdots & &\vdots &\vdots &
 \end{pmatrix}, 
\end{equation}
\end{widetext}
where $\ell, p, m$, and $q$ are the lowest-order modes to be investigated. 
For example, if a particular simulation required the investigation of $-1 \leq \ell \leq 1$ and $0 \leq p \leq 2$, then we would begin incrementing from $\ell = m = -1$ and $ p = q =0$ .
\end{document}